\documentclass[10pt]{article}
\usepackage{amssymb, amsmath, verbatim, epsfig,setspace,mathtools}
\usepackage{subcaption}

\usepackage{mathtools}
\usepackage{xcolor}
\usepackage{amsmath,amssymb,graphicx,epstopdf,bm,bbm}
\usepackage{longtable}
\usepackage{booktabs}
\usepackage{array}
\usepackage{pdflscape}
\usepackage[super,compress]{natbib}
\newcolumntype{L}[1]{>{\raggedright\arraybackslash\hspace{0pt}}p{#1}}

\usepackage{lipsum}  
\title{\bf 
\vspace{-1in}
\LARGE{Decoupling size from magnetism: A length-scale boundary for curvature control in micrometer FePt Janus particles}}
\date{\vspace{-3ex}} 
\author{Natalia Gonzalez-Vazquez$^{1,2,*}$, Eylül Suadiye$^{2}$, Eberhard Goering$^{3}$, Ruben O. Miranda-Rosales$^{1}$,\\ Hilda David$^{2}$, Frank Thiele$^{2}$, Julia Unangst$^{2}$, Andrew K. Schulz$^{4,*}$, Gunther Richter$^{1,2,*}$\\
\small{$^1$Institute for Materials Science, University of Stuttgart, Stuttgart, Germany}\\
\small{$^2$Materials Central Scientific Facility, Max Planck Institute for Intelligent Systems (MPI-IS), Stuttgart, Germany}\\
\small{$^{3}$Max Planck Institute for Solid State Research, Stuttgart, Germany}\\
\small{$^{4}$MPI-IS, Stuttgart, Germany}\\
}
\usepackage[left=1.5cm,right=1.5cm,top=2cm,bottom=2cm]{geometry}
\usepackage{hyperref}

\begin{document}
\maketitle
\noindent{\bf Corresponding authors:} \\
Natalia Gonzalez-Vazquez, Andrew K. Schulz, Gunther Richter\\
Heisenbergstraße 3, 70569 Stuttgart, Germany\\
vazquez@is.mpg.de, aschulz@is.mpg.de, richter@is.mpg.de \\\\

design rules; micromagnetic simulations; FePt

\begin{abstract}
\doublespacing
Curvature reshapes magnetization when a structure's dimensions approach intrinsic magnetic length scales, and much of the current understanding of curvilinear magnetism has been developed from nanoscale shells, wires, and tubes. However, functional magnetic colloids and microrobots are often micrometers in size, where the radius of curvature exceeds these length scales by several orders of magnitude. Whether particle diameter remains an effective parameter for tuning magnetic response in this regime is not well established. We synthesized partially ordered FePt Janus caps on spherical SiO$_2$ particles with diameters of 3--10~$\mu$m, characterized their structure and magnetic response, and extended the investigated range to 1--20~$\mu$m using micromagnetic simulations. Across this range, coercivity, remanence, and hysteresis-loop shape showed no systematic dependence on particle diameter in either experiment or simulation. A dimensionless comparison between the exchange length and radius of curvature ($\ell_{\mathrm{ex}}/R \sim 10^{-3}$--$10^{-4}$) places these particles in a locally planar regime and, when compared with the curvilinear-magnetism literature, identifies the length-scale range beyond which diameter-dependent curvature effects become weak. Size and magnetic response are therefore effectively decoupled within the investigated regime: particle diameter can be selected according to transport, payload, and biocompatibility requirements without introducing a measurable magnetic penalty, but it does not provide an effective route for tuning magnetization reversal. Instead, the magnetic response is governed primarily by material state, including the balance between magnetically hard L1$_0$ and soft A1 FePt, with additional modulation by processing-induced morphology. The resulting length-scale map identifies the regime in which this decoupling is expected to hold and the smaller-length-scale range in which diameter-dependent curvature effects may become significant.
\end{abstract}

\doublespacing

\section*{Introduction}

Magnetic microparticles are engineered around a small set of structural knobs (particle size, shape, coating, and material) each expected to trade against magnetic performance. Among these, particle diameter is rarely a free variable. It is fixed by the operating environment: diameter sets sedimentation and diffusion, hydrodynamic drag and transport, and the magnetic payload a particle can carry, and in the biomedical settings that motivate magnetic colloids, Janus particles, and microrobots it is further constrained by cellular uptake, circulation and clearance, and the vessel and tissue dimensions the particle must navigate~\cite{walther_janus_2013,bozuyuk_high-performance_2022,su_janus_2019,fusco_shape-switching_2015,zhou_system_2023,chen_recent_2017}. Diameter is therefore largely spent on function before magnetic performance is even considered. Rational magnetic design then turns on a specific question: within the narrow size window that an application fixes, does diameter still provide a usable magnetic handle, or must magnetic response be controlled through a different, size-independent axis?

This question is motivated by curvilinear magnetism, where curvature reshapes magnetic behavior once a structure's dimensions approach intrinsic magnetic length scales. In nanoscale magnetic shells, wires, and tubes, curvature can modify domain-wall structure~\cite{landeros_reversal_2007,skoric_domain_2022,fullerton_understanding_2024}, alter switching fields and reversal pathways~\cite{streubel_equilibrium_2012,volkov_experimental_2019,sanz-hernandez_artificial_2020}, and generate curvature-induced exchange and chiral contributions that stabilize nontrivial spin textures~\cite{gaididei_curvature_2014,sheka_micromagnetic_2020,difratta_micromagnetics_2020,streubel_magnetism_2016,sheka_curvature_2015,bran_novel_2022,otalora_curvature-induced_2016,mimica-figari_dzyaloshinskii-moriya_2025}. In spherical caps and shells, reducing the radius can stabilize single-domain, vortex, onion, or skyrmion-like configurations and modify magnetization reversal~\cite{streubel_equilibrium_2012,albrecht_magnetic_2005,ulbrich_magnetization_2006,abdelgawad_magnetic_2018,kravchuk_topologically_2016}; in nanotubes, diameter and wall thickness determine transitions between coherent, transverse, and vortex-like reversal modes~\cite{landeros_reversal_2007,escrig_crossover_2008,bachmann_size_2009,albrecht_experimental_2011}. These studies establish geometry as a genuine control parameter at the nanoscale.

The magnetic particles that actually reach applications, however, are micrometers in diameter, where the radius of curvature exceeds intrinsic magnetic length scales by three to four orders of magnitude. Whether the nanoscale intuition (that shrinking or reshaping the particle tunes its reversal) survives into this regime is not resolved. The relevant comparison is between the radius of curvature, $R = d/2$, and the exchange length,

\begin{equation}
\ell_{\mathrm{ex}} =
\sqrt{\frac{2A}{\mu_0 M_s^2}},
\end{equation}

the characteristic distance over which exchange interactions keep spins aligned. Curvature contributes appreciably to the micromagnetic energy only when $R$ approaches $\ell_{\mathrm{ex}}$~\cite{gaididei_curvature_2014,difratta_micromagnetics_2020}, which has motivated dimensionless descriptors such as the thickness-to-radius ratio ($t/R$) for curved systems~\cite{streubel_magnetism_2016,sheka_micromagnetic_2020,bran_novel_2022}. For chemically ordered L1$_0$ FePt, $\ell_{\mathrm{ex}} \approx 3$--$5$~nm, whereas the particle radii investigated here extend from 0.5 to 10~$\mu$m, giving $\ell_{\mathrm{ex}}/R \sim 10^{-3}$--$10^{-4}$, far from the regime in which curvature-induced effects are most pronounced. Micrometer-scale magnetic Janus particles therefore provide a useful system for testing whether curvature remains a diameter-dependent magnetic parameter when $R \gg \ell_{\mathrm{ex}}$.

FePt Janus caps are particularly suitable for this comparison because geometry and material state provide two distinct magnetic parameters. The same 60~nm FePt coating on micrometer-scale SiO$_2$ particles has previously been shown to be biocompatible \textit{in vitro}, with FePt-coated microparticles retaining more than 85\% macrophage viability and more than 90\% endothelial-cell viability. In the same study, the particles could be detected using both magnetic resonance and photoacoustic imaging without additional contrast agents~\cite{bozuyuk_high-performance_2022}. These results establish FePt-capped microparticles as a realistic and imageable platform for biomedical applications and motivate resolving their magnetic behavior within the application-relevant micrometer size range.

Geometry is defined by the spherical substrate: the FePt-coated hemispherical cap, referred to throughout as the FePt cap, imposes an approximately radial distribution of local easy axes at every investigated diameter. Material state is controlled separately through chemical ordering. The ordered L1$_0$ phase is magnetically hard ($K_u \approx 6.6 \times 10^6$~J/m$^3$), thermally stable, and strongly uniaxial~\cite{weller_high_2000,gutfleisch_fept_2005,weller_review_2016,iwama_structure_2016,sun_monodisperse_2000,pei_direct_2020,sasaki_thermal_2023}, whereas the disordered A1 phase is magnetically softer. Thermal processing can additionally modify cap continuity, grain structure, and local exchange connectivity. Importantly, the radial anisotropy \emph{configuration} imposed by curvature is not equivalent to diameter-dependent \emph{tunability}: geometry may define the anisotropy landscape while changes in particle diameter produce little additional modification of magnetization reversal. Throughout this work, we therefore use the term curvature-dependent magnetization reversal specifically to mean systematic diameter-dependent changes in magnetic response, rather than the existence of the radial anisotropy configuration itself.

Here we test whether particle size tunes magnetization reversal in the micrometer regime, and identify what does. FePt caps of constant nominal thickness were deposited on spherical SiO$_2$ particles with diameters of 3--10~$\mu$m and characterized structurally and magnetically, while micromagnetic simulations extended the investigated range to 1--20~$\mu$m and allowed geometry, anisotropy configuration, and phase fraction to be varied independently. Across this range, we find no systematic dependence of coercivity, remanence, or hysteresis-loop shape on particle diameter in either experiment or simulation. In contrast, the relative contributions of magnetically hard L1$_0$ and soft A1 FePt strongly modify the simulated magnetic response, while processing-induced morphology provides a plausible additional source of magnetic softening. By placing these results within a literature-based length-scale map, we identify a micrometer-scale regime in which curvature defines the anisotropy configuration but particle diameter no longer provides an effective parameter for tuning magnetization reversal.

\section*{Results and Discussion}

To place the investigated system within the broader field of curvilinear magnetism, we compared literature-reported curved ferromagnetic systems using the dimensionless ratio $\ell_{\mathrm{ex}}/R$ (Figure~\ref{fig:RegimeMap}b; Table~S7). The compiled systems span spherical caps and shells, nanotubes, rolled magnetic membranes, curved nanowires, and helical structures across radii ranging from the nanometer to the micrometer scale. Many systems exhibiting pronounced curvature-dependent magnetic behavior occupy regimes in which the radius of curvature approaches intrinsic magnetic length scales. By comparison, the FePt Janus caps investigated here lie at much smaller values of $\ell_{\mathrm{ex}}/R$, providing an experimental platform for testing whether diameter remains an effective magnetic design parameter when $R \gg \ell_{\mathrm{ex}}$. The literature comparison also shows that the presence of magnetic effects at large radii does not necessarily imply a diameter-dependent curvature response. Rolled membranes, curved conduits, spirals, and three-dimensional helices can exhibit modified anisotropy, domain-wall pinning, spin-wave behavior, or chirality even when $\ell_{\mathrm{ex}}/R$ is small~\cite{muller_tuning_2009,balhorn_spin-wave_2010,lewis_magnetic_2009,brajuskovic_understanding_2021,sanz-hernandez_artificial_2020}. In these systems, geometry changes the global magnetic boundary conditions, topology, strain state, or propagation pathway. By contrast, the present study varies the radius of an otherwise geometrically similar hemispherical cap while maintaining nominal film thickness and material composition. This distinction allows us to test whether geometric scaling alone modifies magnetization reversal within a fixed class of curved magnetic structures.

This distinction is particularly relevant for biomedical and microrobotic applications, where particle diameter cannot be selected solely according to magnetic performance (Figure~\ref{fig:RegimeMap}a). Micrometer-scale magnetic Janus particles must also satisfy requirements related to transport through complex fluids, hydrodynamic drag, magnetic payload, controllability, cellular interactions, and safety. If magnetization reversal remains approximately invariant across an application-relevant size window, particle diameter can be selected according to these transport and biological requirements without necessarily introducing a substantial magnetic penalty. Conversely, when distinct magnetic responses are required, for example for future selective actuation, sorting, staged delivery, or swarm-level control, the present results suggest that phase ordering and microstructural design are more promising control parameters than particle diameter (Figure~\ref{fig:RegimeMap}c). We note that these applications motivate the present design question but are not demonstrated here.

To determine how geometric curvature, phase ordering, and morphology contribute to magnetization reversal, we combined FePt Janus particle fabrication, structural characterization, SQUID magnetometry, and micromagnetic simulations within a single experimental-computational framework. As summarized in Figure~\ref{fig:Workflow}, this approach enabled direct comparison between ensemble-averaged measurements and idealized single-cap simulations, allowing the influence of particle diameter to be separated from changes associated with anisotropy configuration, phase-ordering state, and processing-induced structural evolution~\cite{jia_propulsion_2025}.

A central challenge in curved magnetic systems is that geometry can influence the magnetic response through more than one mechanism. The hemispherical geometry defines the local surface-normal distribution and therefore configures the anisotropy landscape of the FePt cap. At the same time, changing particle diameter modifies the radius of curvature and may, in principle, alter coercivity, remanence, or magnetization-reversal pathways. These two effects are not equivalent. Throughout this work, ``curvature-dependent reversal'' refers specifically to systematic changes in magnetic response with particle diameter, whereas the curvature-imposed radial anisotropy configuration is treated as a geometric characteristic shared by all investigated caps.

The experimental design therefore maintained nominal film thickness, composition, and processing conditions as constant as possible while varying particle diameter from 3 to 10~$\mu$m. Micromagnetic simulations extended this range to 1--20~$\mu$m and provided a controlled reference in which geometric and material parameters could be modified independently. This combination allowed three contributions to be examined separately: the effect of particle diameter, the effect of radial compared with uniform anisotropy, and the effect of varying the contribution of magnetically hard L1$_0$ and soft A1 FePt.

The following sections first establish the structural and compositional consistency of the FePt Janus caps, followed by experimental evaluation of diameter-dependent magnetic behavior. We then use micromagnetic simulations to distinguish the configurational role of radial anisotropy from the effect of particle diameter and finally examine how phase ordering and processing-induced morphology modify the magnetic response.

\subsection*{Fabrication and Structural Characterization of FePt Janus Caps Across Curvatures}

\begin{figure}[thbp]
      \centering
\includegraphics[width=1\textwidth,page=1,trim=0.0in 0in 0in 0.0in,clip]{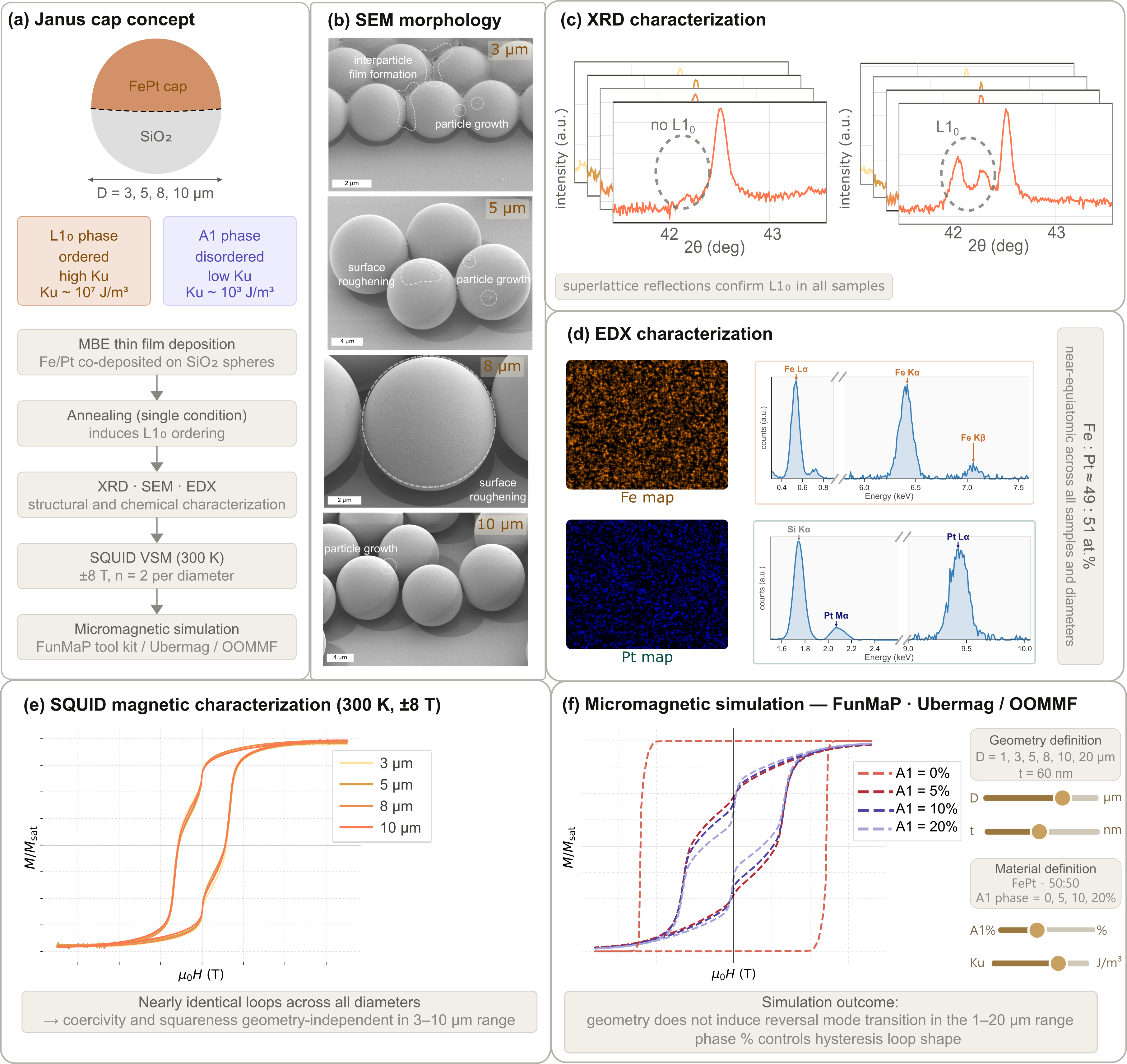}
    \caption{\textbf{Integrated experimental–computational workflow for FePt Janus caps across curvature.}
    (a) Schematic of FePt-coated SiO$_2$ Janus particles with diameters of 3--10~$\mu$m, illustrating the coexistence of magnetically hard L1$_0$ and soft A1 phases and the overall workflow from fabrication to simulation.
    (b) Representative SEM images showing continuous FePt caps across all particle diameters.
    (c) X-ray diffraction (XRD) patterns indicating partial L1$_0$ ordering with no diameter-dependent structural variation.
    (d) Energy-dispersive X-ray spectroscopy (EDX) confirming homogeneous elemental distribution.
    (e) SQUID hysteresis loops demonstrating consistent magnetic behavior across particle sizes.
    (f) Micromagnetic simulations of ideal FePt caps used to isolate intrinsic curvature effects.}\label{fig:Workflow}
\end{figure}

To probe diameter-dependent curvature effects, monodisperse SiO$_2$ spheres with diameters of 3, 5, 8, and 10~$\mu$m were assembled into near-monolayers \textit{via} drop-casting of diluted 5\% (w/v) suspensions. The dilution factor was adjusted according to geometric packing considerations to obtain comparable surface coverage across particle sizes~\cite{ohring_chapter_2002}. All samples were coated with FePt under the same deposition conditions and with a constant nominal thickness of 60~nm. Particle diameter was therefore the primary intentionally varied geometric parameter in the experimental series (Figure~\ref{fig:Workflow}a).

\begin{figure}[thbp]
      \centering
\includegraphics[width=1\textwidth,page=1,trim=0.0in 0in 0in 0.0in,clip]{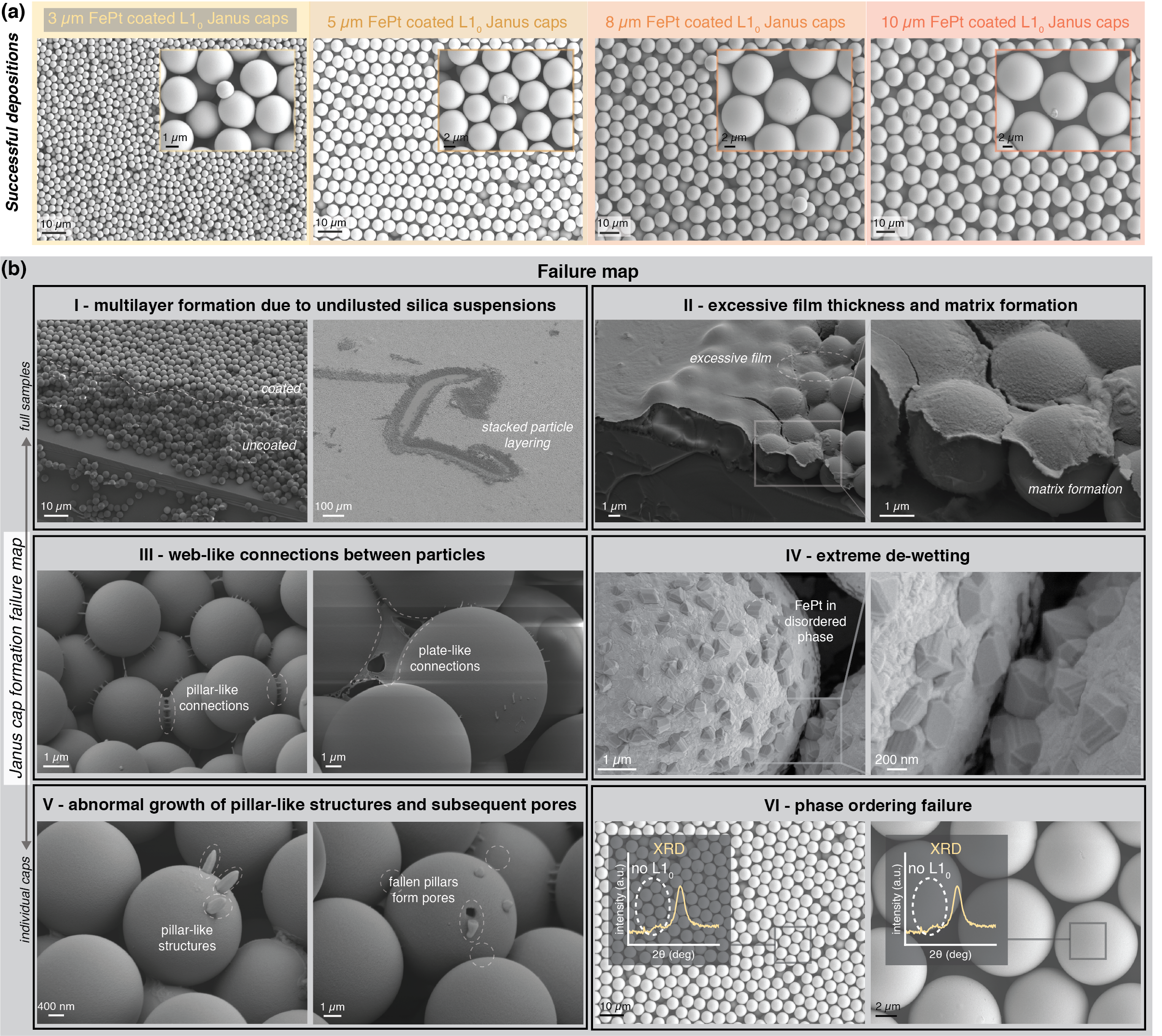}
    \caption{\textbf{Processing window linking FePt cap morphology and phase ordering.}
    (a) Representative SEM images of successfully fabricated FePt-coated SiO$_2$ particle monolayers across diameters (3--10~$\mu$m), demonstrating uniform coverage and reproducible cap formation. Insets highlight local ordering and surface morphology at higher magnification.
    (b) Failure map summarizing dominant morphological instabilities observed during process development. (I) Multilayer formation due to insufficient dilution, which leads to shadowing and inhomogeneous coatings. (II) Excessive film thickness resulting in matrix formation and loss of particle individuality. (III) Interparticle connections forming web-like or plate-like structures due to lateral diffusion and coalescence. (IV) Extreme dewetting forming FePt in the disordered phase on the particles. (V) Abnormal pillar-like growth and pore formation associated with limited adatom mobility. (VI) Phase-ordering failure, characterized by incomplete L1$_0$ transformation as demonstrated by inset XRD spectra. The failure map is ordered from full samples (I-II) to individual caps (V-VI).}\label{fig:Map}
\end{figure}

SEM imaging confirmed near-monolayer coverage with predominantly hexagonal particle packing across all investigated diameters, while higher-magnification images showed largely continuous FePt cap formation (Figures~\ref{fig:Workflow}b and~\ref{fig:Map}a). No systematic differences in packing arrangement or multilayer formation were observed as a function of particle size. The process window summarized in Figure~\ref{fig:Map}b documents the principal fabrication outcomes encountered during process optimization. Insufficient dilution, excessive film thickness, variations in deposition conditions, and thermal processing produced multilayer formation, matrix-like coatings, abnormal pillar growth, interparticle coalescence, extensive dewetting, or incomplete phase ordering. These conditions were excluded from the magnetic dataset but defined the processing boundaries required to obtain largely continuous FePt caps with partial L1$_0$ ordering. The processing map further illustrates why phase ordering and morphology must be considered together: the thermal treatment required to promote chemical ordering simultaneously activates diffusion-driven structural evolution, producing outcomes ranging from incomplete ordering to extensive dewetting (Figure~\ref{fig:Map}b)~\cite{thompson_solid-state_2012,jiran_capillary_1990}.

Weak features near the expected positions of the (001)$_{\mathrm{fct}}$ and (110)$_{\mathrm{fct}}$ superlattice reflections were observed around $23.2^\circ$ and $41.2^\circ$, respectively, consistent with the onset of L1$_0$ chemical ordering (Figure~\ref{fig:Workflow}c). Fundamental reflections, including the overlapping (111)$_{\mathrm{fcc}}$/(111)$_{\mathrm{fct}}$ contribution and the (002)$_{\mathrm{fct}}$ reflection, remained dominant~\cite{speliotis_microstructure_2015}. Given the thin, partial-coverage geometry of the FePt caps and the strong substrate contribution, the XRD data were interpreted qualitatively rather than used to determine a quantitative L1$_0$ order parameter or A1 phase fraction, consistent with previous reports on curved FePt-coated microparticle systems~\cite{bozuyuk_high-performance_2022}. We therefore refer to the annealed samples as partially ordered FePt caps without assigning a specific L1$_0$-A1 phase fraction.

Energy-dispersive X-ray spectroscopy (EDX) confirmed near-equiatomic composition across all samples (Fe:Pt $\approx$ 49:51 at.\%), with no detectable systematic variation as a function of particle diameter (Figure~\ref{fig:Workflow}d). Together, the XRD and EDX results showed no systematic diameter-dependent variation in the measured phase signature or elemental composition within the sensitivity of the applied methods. The subsequent magnetic comparison could therefore focus on whether changing particle diameter modified the hysteresis response under otherwise comparable processing conditions.

Magnetic measurements (Figure~\ref{fig:Workflow}e) showed closely overlapping hysteresis responses across the investigated particle diameters. The following section evaluates these differences quantitatively and compares the experimental behavior with micromagnetic simulations of idealized FePt caps (Figure~\ref{fig:Workflow}f).

\subsection*{Diameter-Independent Magnetization Reversal}

The magnetic response was measured using SQUID magnetometry. As shown in Figure~\ref{fig:SQUID}a, the normalized hysteresis loops ($M/M_s$) for all particle diameters converged onto a nearly identical curve, exhibiting sharp switching and high remanence. This invariance is quantified in Table~\ref{tab:SQUID_sim_comparison}. 

Across the four particle diameters, repeated SQUID measurements yielded coercivity values within a narrow range of $\mu_0 H_c = 1.03$--$1.21$~T. The variation between repeated measurements of the same sample was comparable to the differences observed between particle diameters. The remanence ratio was similarly consistent, ranging from $M_r/M_s = 0.917$ to $0.976$, while hysteresis-loss values ranged from 89.0 to 110.2~J/kg. The pooled coercivity across all measurements was $\mu_0 H_c = 1.13$~T, and no systematic monotonic or non-monotonic dependence on particle diameter was resolved. Figure~\ref{fig:SQUID}b therefore shows that the observed variations remain within the measurement-to-measurement scatter of the investigated series.

Taken together, the experimental SQUID data show no measurable diameter-dependent influence on magnetization reversal across the investigated micrometer-scale curvature range. This result shows that the expectation derived from nanoscale curvilinear systems cannot be directly extended to micrometer-scale FePt Janus caps~\cite{gaididei_curvature_2014,streubel_magnetism_2016,mimica-figari_dzyaloshinskii-moriya_2025}. For comparison, single-shot measurements on as-deposited (A1-phase) FePt caps yielded $\mu_0H_c \approx 3-8$~mT across the same diameter range (Table~S3), more than two orders of magnitude below the annealed values. This strong contrast supports the conclusion that the high coercivity of the annealed caps arises primarily from ordering-related changes rather than from particle diameter or from the intrinsic response of as-deposited FePt at this thickness and geometry.

We emphasize that this diameter independence refers to the \textit{direct} magnetic response at a fixed material state: once the radial anisotropy configuration is established, magnetization reversal does not scale with the particle radius because $R \gg \ell_{\mathrm{ex}}$. It does not exclude an \textit{indirect} route in which particle size influences the phase-ordering and morphological state actually achieved during processing on a curved substrate: a distinct, kinetics-driven effect that we examine separately below and revisit as an outlook in the Conclusion.

\subsection*{Morphological Evolution of Curved FePt Caps}

Scanning electron microscopy (SEM) analysis after annealing showed that the FePt caps remained largely continuous across all particle diameters while developing surface roughness and rim-associated morphological features (Figure~\ref{fig:Morphology}a). This morphology differs from the complete fragmentation often observed in planar thin films and instead corresponds to early-stage solid-state dewetting \cite{thompson_solid-state_2012, mullins_theory_1957, srolovitz_capillary_1986}.

As illustrated in Figure~\ref{fig:Morphology}c, the hemispherical geometry imposed a thickness gradient ($t(\theta) = t_0 \cos\theta$, with $t_0 \approx 60$~nm at the pole), leading to a continuous reduction in thickness toward the cap perimeter. Consequently, the local thickness approached zero near the particle's equator, where diffusion-driven evolution is fastest.

Consistent with this thickness dependence, we found morphological changes were localized at the rim, while the central cap region remained comparatively smooth and continuous. The presence of interparticle bridges in the 3~$\mu$m samples (Figure~\ref{fig:Morphology}b) further indicated enhanced lateral mass transport at small interparticle separations. However, despite these structural variations, no corresponding change in magnetic switching behavior was observed.

\subsection*{Micromagnetic Simulations: Intrinsic Baseline}

Micromagnetic simulations of ideal FePt caps provided a reference for intrinsic behavior using the workflow summarized in Figure~\ref{fig:CurvatureIndependency}a. The simulations are intended as an idealized, anisotropy-dominated baseline rather than as a fully microstructure-resolved model of the experimental caps. In particular, the model isolates the effects of cap diameter, radial anisotropy configuration, and prescribed A1/L1$_0$ phase fraction under otherwise controlled conditions, with exchange, anisotropy, Zeeman, and magnetostatic contributions included in the micromagnetic energy functional (Figure~\ref{fig:CurvatureIndependency}b). It does not explicitly resolve the experimental grain-boundary network, rim roughness, local dewetting morphology, or domain-wall structure at the exchange-length scale. The simulated trends should therefore be interpreted as a controlled comparison of design parameters, showing that diameter variation alone does not produce a systematic reversal transition in the locally planar micrometer regime, whereas changes in anisotropy configuration and phase-ordering state strongly modify the hysteresis response. As shown in Figure~\ref{fig:CurvatureIndependency}c, simulated hysteresis loops remained effectively invariant across all particle diameters from 1 to 20~$\mu$m. This behavior differs from the size-dependent transitions reported for nanoscale caps, shells, and nanotubes, where changes in radius can alter the balance between exchange, anisotropy, and magnetostatic energies and thereby stabilize different reversal modes or magnetic states~\cite{streubel_equilibrium_2012,amaladass_size_2007,brandt_size-dependent_2013,escrig_crossover_2008,bachmann_size_2009}. In the present system, however, all investigated radii remain substantially larger than the FePt exchange length, and the hemispherical geometry remains self-similar across the diameter series.

This is reflected quantitatively in Figure~\ref{fig:CurvatureIndependency}d and Table~\ref{tab:SQUID_sim_comparison}: at every fixed phase composition, simulated coercivity varies by less than 5\% across the full 1--20~$\mu$m diameter range. For fully ordered L1$_0$ FePt, corresponding to an A1 fraction of 0\%, $\mu_0 H_{\mathrm{c}}$ spans 11.85~T (1~$\mu$m), 12.09~T (3~$\mu$m), 12.17~T (5~$\mu$m), 12.16~T (8~$\mu$m), 12.36~T (10~$\mu$m), and 12.35~T (20~$\mu$m), corresponding to a mean of $\mu_0 H_c = 12.17 \pm 0.18$~T (relative spread 4.2\%). At the representative case shown in Figure~\ref{fig:CurvatureIndependency}c, corresponding to an A1 fraction of 5\%, the mean coercivity across six diameters is $\mu_0 H_c = 5.64 \pm 0.11$~T (spread $\sim$2\%) with mean remanence ratio $M_r/M_s = 0.467 \pm 0.010$ ($\sim$2\%). These residual variations remain within the numerical resolution of the simulations and do not indicate a systematic curvature-dependent trend.

A subtle but important distinction emerges from the simulations. Curvature did play a role in the magnetic response, but only by determining the \textit{spatial distribution} of the easy axis: in a hemispherical FePt cap, the easy axis followed the local surface normal, $\hat{e}_u(\mathbf{r}) \parallel \hat{r}$, producing a distributed (radial) anisotropy configuration that differed qualitatively from the uniform configuration of a planar film.

As shown in Figure~\ref{fig:PhaseDependency}a, simulations with a uniaxial easy axis ($\hat{e}_u \parallel \hat{z}$) at an A1 fraction of 5\% yielded a square loop with $\mu_0 H_c = 11.62$~T, while simulations with a radial easy axis on the same geometry and phase composition yielded a sheared loop with $\mu_0 H_c = 5.53$~T, a $2.1\times$ reduction in coercivity arising solely from the change in easy-axis distribution. The radial configuration produced a more strongly sheared hysteresis response than the uniform uniaxial configuration, supporting the conclusion that the curvature-imposed easy-axis distribution contributes to the experimental reversal behavior.

These comparisons showed that anisotropy configuration and phase disorder provide distinct but comparably strong contributions to the simulated magnetic response. At a fixed A1 fraction of 5\%, changing from a uniform to a radial easy-axis distribution reduced $\mu_0 H_c$ by a factor of approximately 2.1. Within the radial configuration, increasing the A1 fraction from 0\% to 5\% produced a comparable reduction of approximately 2.2. Geometry therefore established the anisotropy configuration, while phase ordering provided an additional route for tuning the response within that configuration. What curvature did not do is produce a diameter-dependent reversal once the radial configuration was established. Because $R \gg \ell_{\mathrm{ex}}$ ($\ell_{\mathrm{ex}}/R \sim 10^{-3}$--$10^{-4}$), the local easy-axis distribution is geometrically similar across diameters, and no significant diameter scaling emerged within the investigated regime. Therefore, curvature acted as a \textit{configurational} rather than a \textit{tunable} parameter in this regime: it defined the anisotropy distribution, but did not allow that distribution to be modified through diameter selection.

\subsection*{Material State, Not Diameter, as the Tuning Axis}

In contrast to particle diameter, simulations showed a strong and systematic dependence on phase composition. The prescribed A1 fractions were introduced as spatially distributed low-anisotropy regions within the harder L1$_0$ matrix, as illustrated in Figure~\ref{fig:PhaseDependency}b. As summarized in Figure~\ref{fig:PhaseDependency}c--d and Table~\ref{tab:SQUID_sim_comparison}, increasing the A1 fraction from 0\% to 20\% reduced the coercive field from $\mu_0 H_{\mathrm{c}} = 12.17 \pm 0.18$~T at an A1 fraction of 0\% to $5.64 \pm 0.11$~T at 5\%, $5.23 \pm 0.10$~T at 10\%, and $3.81 \pm 0.10$~T at 20\%. The remanence ratio dropped in parallel from $M_r/M_s \approx 1.0$ (0\% A1) to $0.47 \pm 0.01$ (5\% A1), $0.41 \pm 0.01$ (10\% A1), and $0.33 \pm 0.01$ (20\% A1). The most pronounced change occurs at low A1 fractions: introducing 5\% disorder reduced $\mu_0 H_{\mathrm{c}}$ by a factor of 2.2 ($12.2 \rightarrow 5.6$~T) and $M_r/M_s$ by approximately 50\% ($1.0 \rightarrow 0.47$). Increasing A1 fraction further (5\% $\rightarrow$ 20\%) reduced $\mu_0 H_{\mathrm{c}}$ by an additional factor of 1.5 ($5.6 \rightarrow 3.8$~T), which confirmed that even small deviations from full L1$_0$ order strongly modify hysteresis behavior.

The experimental coercivity ($\mu_0 H_c = 1.13$~T pooled mean, range $1.03$--$1.21$~T) was approximately $11\times$ smaller than the fully ordered limit ($12.17 \pm 0.18$~T) and $\sim$3.4$\times$ smaller than even the most disordered simulated case (20\% A1, $3.81 \pm 0.10$~T). This large coercivity deficit was not, however, accompanied by a corresponding loss of remanence: the measured remanence ratio ($M_r/M_s = 0.917$--$0.976$) remained close to the fully ordered simulated limit, far above the value the phase-fraction model predicts for a state soft enough to reproduce the measured coercivity ($M_r/M_s \approx 0.47$ at 5\% A1). Figure~\ref{fig:PhaseDependency}d makes this decoupling explicit: overlaying the measured coercivity and remanence on the simulated phase-fraction trend shows that the measured coercivity lies a factor of $\sim$3.4 below even the most disordered simulated case while the measured remanence sits near the ordered end. A single bulk A1 fraction cannot reproduce both observations simultaneously, so the samples are therefore better described as retaining an ordered-like remanent state while being strongly microstructurally softened, rather than as being represented by a single homogeneous A1 phase fraction. The idealized two-phase model therefore provides an upper bound on coercivity, and the residual gap may arise from experimental microstructural features that are not represented by the prescribed phase-fraction model, including partial dewetting, particle and grain formation, and rim and surface roughening, all of which lower the local reversal barrier while preserving the aligned remanent state. Hysteresis loss is omitted from this comparison because the measured and simulated loops were swept over different field ranges ($\pm$7~T experiment versus $\pm$18~T simulation), so their enclosed areas are not directly comparable.

The strong sensitivity of $H_c$ to A1 fraction can be understood from the large anisotropy contrast between the two modeled phases. The disordered A1 phase exhibited a vanishingly small magnetocrystalline anisotropy ($K_{\mathrm{A1}} = 1.0 \times 10^4$~J/m$^3$, approximately three orders of magnitude smaller than $K_{L1_0} \approx 6.6 \times 10^6$~J/m$^3$), and therefore reduces the local effective anisotropy according to
\begin{equation}
K_{\mathrm{eff}} \approx (1 - f) K_{L1_0} + f K_{\mathrm{A1}},
\label{eq:Keff}
\end{equation}
where $f$ is the A1 volume fraction. Inserting $f = 0.05$ predicted a $\sim$5\% reduction in mean anisotropy, yet the simulated coercivity dropped by $\sim$54\%, indicating that the contribution of the A1-like regions cannot be described solely by a linear volume average of the anisotropy. 

This nonlinear response indicates that the A1-like regions do not act solely through a volume-averaged reduction in anisotropy. Instead, their low local anisotropy provides energetically favorable regions for early magnetization rotation, while exchange coupling to the surrounding L1$_0$ matrix can assist the subsequent reversal of harder regions. This interpretation is consistent with the progressive shearing of the simulated loops and with exchange-spring behavior reported for coupled hard-soft magnetic systems~\cite{kneller_exchange-spring_1991}. The gradual descent from $M_r$ as the field reverses is therefore consistent with early rotation in magnetically softer regions, while the steeper decrease at larger reverse fields reflects the subsequent reversal of the harder L1$_0$ contribution.

\subsection*{Coupled Role of Ordering and Morphology}

At the highest simulated disorder level (20\% A1, $\mu_0 H_c = 3.81 \pm 0.10$~T), the simulated coercivity exceeded the experimental value ($1.13$~T pooled mean, range $1.03$--$1.21$~T) by a factor of $\sim$3.4. This residual gap was not captured by the simulated phase-fraction range alone and is consistent with additional magnetic softening associated with the morphological evolution observed by SEM (Figure~\ref{fig:Morphology}).

Dewetting kinetics follow $\tau_n \propto h^4 / D_s$ \cite{thompson_solid-state_2012, mullins_theory_1957}, where $h$ is the local film thickness and $D_s$ is the surface diffusion coefficient. The fourth-power dependence on thickness implied that the rim region (where $t \rightarrow 0$ as $\theta \rightarrow \pi/2$, see Figure~\ref{fig:Morphology}c) evolved orders of magnitude faster than the apex. Consistent with this, SEM analysis localized morphological changes (roughening, grain-boundary grooving, and partial edge retraction) at the cap perimeter while the central region remains comparatively continuous. These morphological features can influence magnetic behavior through two mechanisms. First, \textbf{exchange decoupling}: grain-boundary grooving may reduce effective exchange coupling when characteristic feature sizes approach or exceed the exchange length, rather than as a direct consequence of structural discontinuity. This distinction is important, as structural fragmentation alone does not necessarily alter hysteresis behavior if exchange connectivity is preserved \cite{herzer_nanocrystalline_1992}. Second, \textbf{localized soft regions}: the rim thickness gradient creates regions where the cap is sufficiently thin for finite-size effects (\textit{e.g.}, reduced anisotropy due to surface contributions) to lower $K_{\mathrm{eff}}$, generating additional nucleation sites for reversal.

Within this framework, curvature does not act as a direct tuning knob. Rather, it influences magnetic behavior indirectly by setting the thickness distribution $t(\theta) = t_0 \cos\theta$, which in turn biases dewetting kinetics and the density of soft nucleation sites. This explains why experimental coercivity fell below the chemically-disordered simulation limit while remaining diameter-invariant: the morphological softening was curvature-driven through $t(\theta)$, but the integrated effect across the cap did not scale with $R$ in the micrometer-scale curvature regime ($R \gg \ell_{\mathrm{ex}}$) with fixed film thickness ($t \approx 60$~nm).

In FePt caps at micrometer-scale curvature (set by particle diameter) but nanometer-scale thickness ($t \approx 60$~nm), three effects must be distinguished. (i) Curvature configures the magnetic system by setting a radial easy-axis distribution, which fundamentally alters loop shape relative to a planar film ($H_c$ reduces from $\sim$12 T uniaxial to $\sim$5.6 T radial). (ii) Chemical ordering tunes the magnitude of the anisotropy contrast within that configuration, scaling $H_c$ inversely with increasing A1 fraction. (iii) Morphology may further modulate the response through changes in exchange connectivity and localized magnetic softening, providing a plausible contribution to the remaining difference between simulation ($\sim3.8$~T at 20\% A1) and experiment ($\sim1.1$~T). What curvature did \textit{not} do is produce a diameter-dependent response in this regime.

The $\ell_{\mathrm{ex}}/R$ criterion is predictive as well as descriptive because it indicates the length-scale range in which diameter-dependent curvature effects are expected to become significant. With $\ell_{\mathrm{ex}} \approx 4$~nm for the present material, the caps studied here ($R \approx 0.5$--$10~\mu$m) sit at $\ell_{\mathrm{ex}}/R \sim 10^{-3}$--$10^{-4}$, deep in the decoupled regime. Curvature-coupled reversal would return only as $\ell_{\mathrm{ex}}/R$ approaches order $10^{-1}$--$1$, corresponding to particle radii of a few to a few tens of nanometers (diameters of roughly $10$--$80$~nm), and a higher-$M_s$ material would shift this boundary to still smaller sizes since $\ell_{\mathrm{ex}} \propto 1/M_s$. This provides a concrete design criterion for other curved magnetic microsystems, for instance sub-100~nm curved films or nanoparticle assemblies, where geometry could be used deliberately as a reversal knob. For the biomedical particle class that motivates this work, however, that regime is not accessible: controllable magnetic actuation and imaging require the micrometer-scale magnetic volume characteristic of the microrobotic and colloidal systems discussed above, and the size window fixed by transport, uptake, and clearance constraints lies one to three orders of magnitude above the crossover. Across the micrometer-scale range relevant to the particle systems considered here, the system is therefore expected to remain within the decoupled regime. Within this range, the results support tuning magnetic response primarily through material state and microstructure rather than through particle diameter.

\section*{Conclusion}

\begin{figure}[thbp]
      \centering
\includegraphics[width=1\textwidth,page=1,trim=0.0in 0in 0in 0.0in,clip]{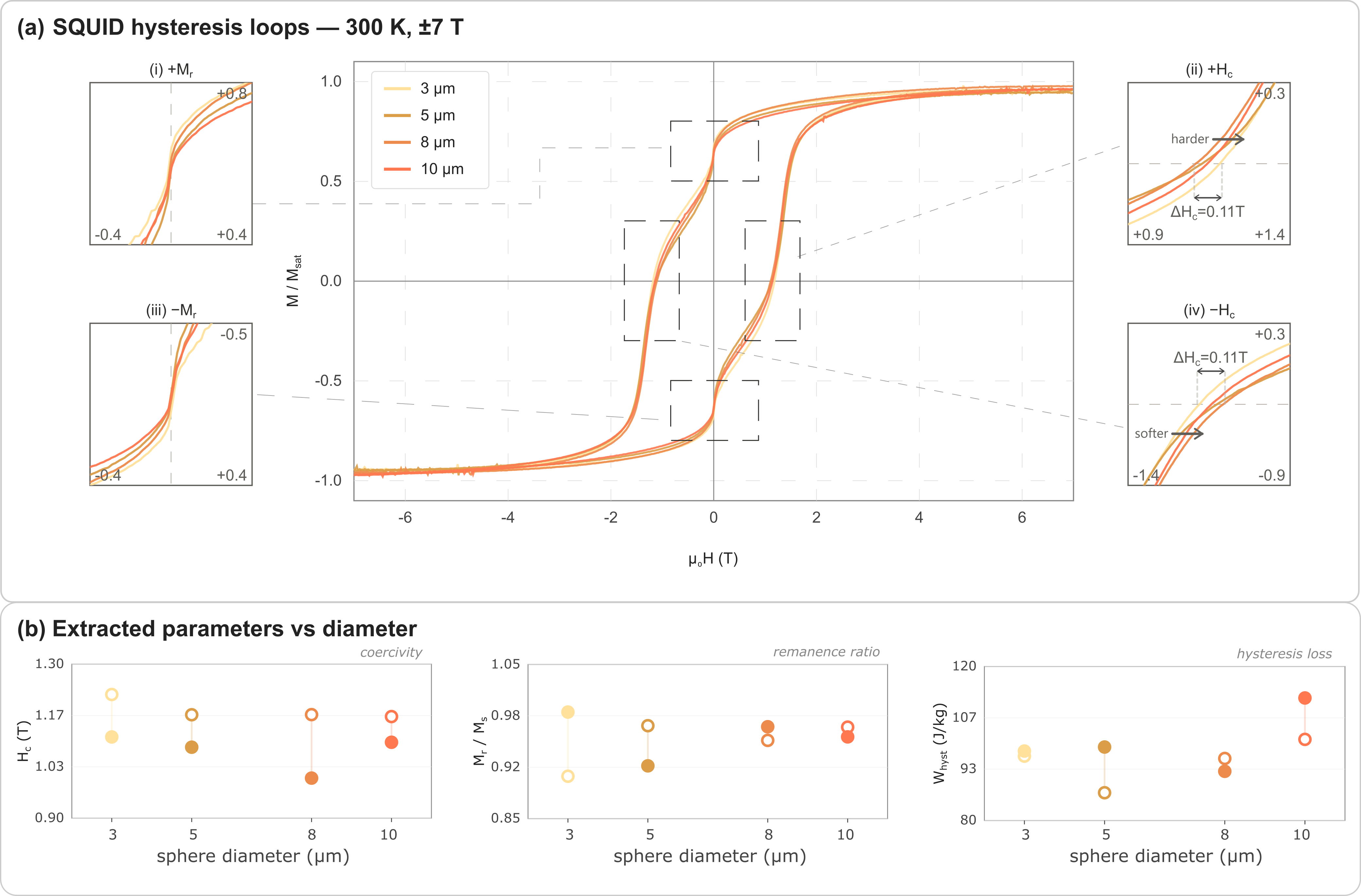}
\caption{
SQUID magnetometry demonstrating diameter-independent magnetization reversal in FePt Janus caps.
(a) Normalized hysteresis loops measured at 300~K ($\pm$7~T) for particles with diameters of 3, 5, 8, and 10~$\mu$m. The loops collapse onto a common profile, indicating nearly identical switching behavior across all sizes. Insets highlight remanent magnetization ($\pm M_r$) and coercive fields ($\pm H_c$), showing only minor variations between samples.
(b) Extracted magnetic parameters as a function of particle diameter: coercivity ($H_c$), remanence ratio ($M_r/M_s$), and hysteresis loss ($W_{\mathrm{hyst}}$). The two markers per diameter correspond to repeated SQUID measurements acquired from the same sample chip; vertical lines connecting paired markers serve only as a visual guide. Individual measurements are shown to convey measurement repeatability and should not be interpreted as independently fabricated sample replicates. Across the investigated series, coercivity, remanence ratio, and hysteresis loss show no systematic dependence on particle diameter. The pooled coercivity is $\mu_0 H_c = 1.13$~T, with an overall range of $1.03$--$1.21$~T.
}\label{fig:SQUID}
\end{figure}

\begin{figure}[thbp]
      \centering
\includegraphics[width=0.95\textwidth,page=1,trim=0.0in 0in 0in 0.0in,clip]{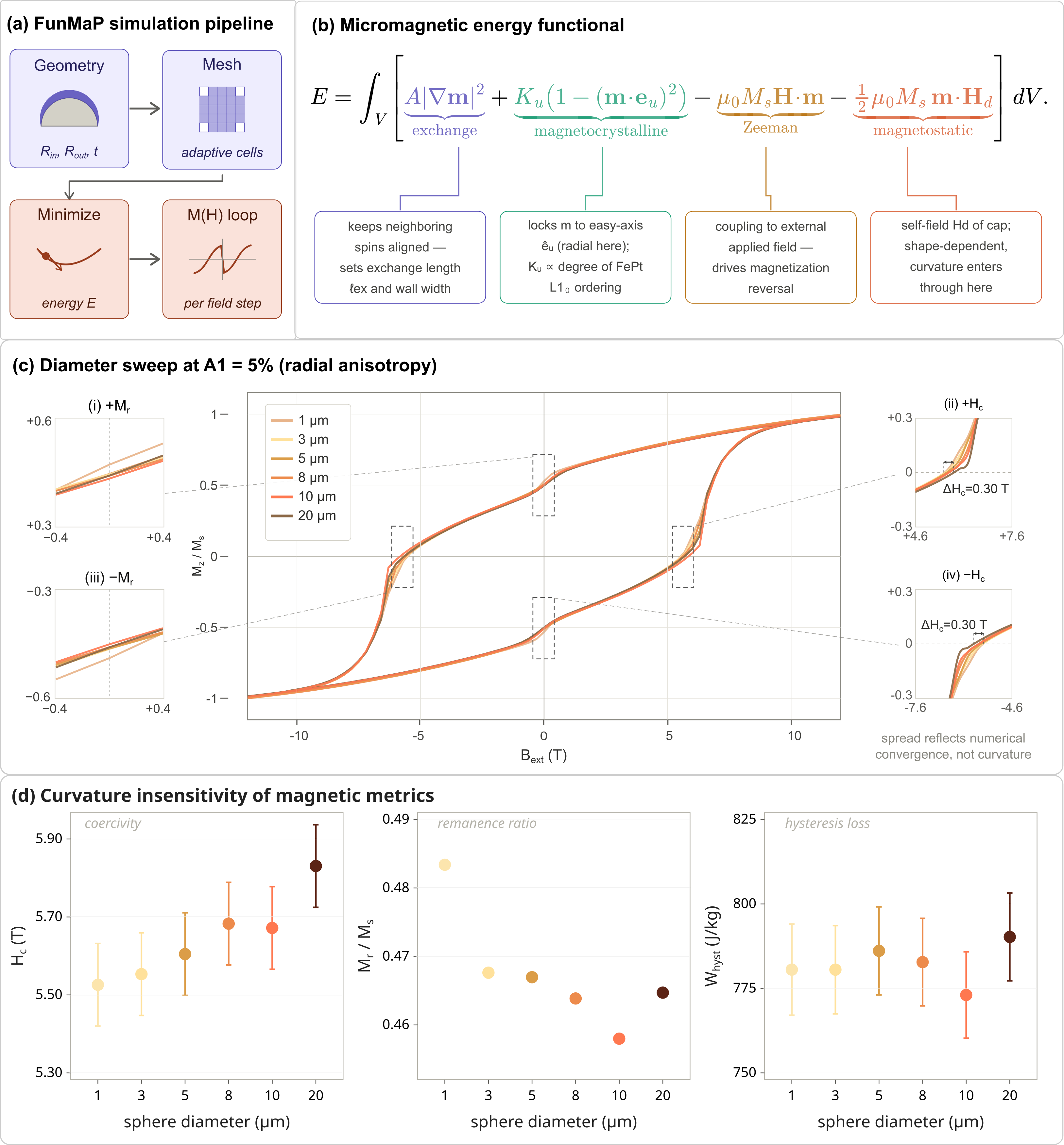}
      \caption{
      Micromagnetic analysis of magnetization reversal in curved FePt caps.
    (a) Simulation workflow, including geometry definition, geometry-dependent discretization, energy minimization, and field-dependent hysteresis calculation.
    (b) Micromagnetic energy functional incorporating exchange, anisotropy, Zeeman, and magnetostatic contributions, with curvature encoded through the shell geometry and radial anisotropy distribution.
    (c) Simulated hysteresis loops for caps with diameters of 1, 3, 5, 8, 10, and 20~$\mu$m at an A1 fraction of 5\% with radial anisotropy, showing nearly identical reversal behavior across all sizes. Across the six diameters, the mean coercivity is $\mu_0 H_c = 5.64 \pm 0.11$~T (spread $\sim$2\%) and the mean remanence ratio is $M_r/M_s = 0.467 \pm 0.010$ (spread $\sim$2.2\%). Insets highlight small variations in coercive field ($\Delta H_{\mathrm{c}} \approx 0.30$~T) that remain within the numerical resolution of the simulations. (d) Extracted coercivity, remanence ratio, and hysteresis loss as a function of particle diameter, showing no systematic diameter-dependent trend.
      }\label{fig:CurvatureIndependency}
\end{figure}

We combined FePt Janus particle fabrication, structural characterization, SQUID magnetometry, and micromagnetic simulations to determine how particle diameter, anisotropy configuration, phase ordering, and morphology contribute to magnetization reversal in curved FePt caps. Across experimentally investigated particle diameters of 3--10~$\mu$m, the measured hysteresis loops showed no systematic diameter-dependent variation in coercivity, remanence, or switching behavior (Figure~\ref{fig:SQUID}; Table~\ref{tab:SQUID_sim_comparison}). Micromagnetic simulations extended this result to diameters of 1--20~$\mu$m and showed that, within the investigated regime, the particle radius remained substantially larger than the FePt exchange length (Figure~\ref{fig:CurvatureIndependency}). The hemispherical geometry therefore configured the spatial distribution of the easy axis, but changing particle diameter did not introduce an additional systematic modification of the magnetic response.

\begin{figure}[thbp]
      \centering
\includegraphics[width=1\textwidth,page=1,trim=0.0in 0in 0in 0.0in,clip]{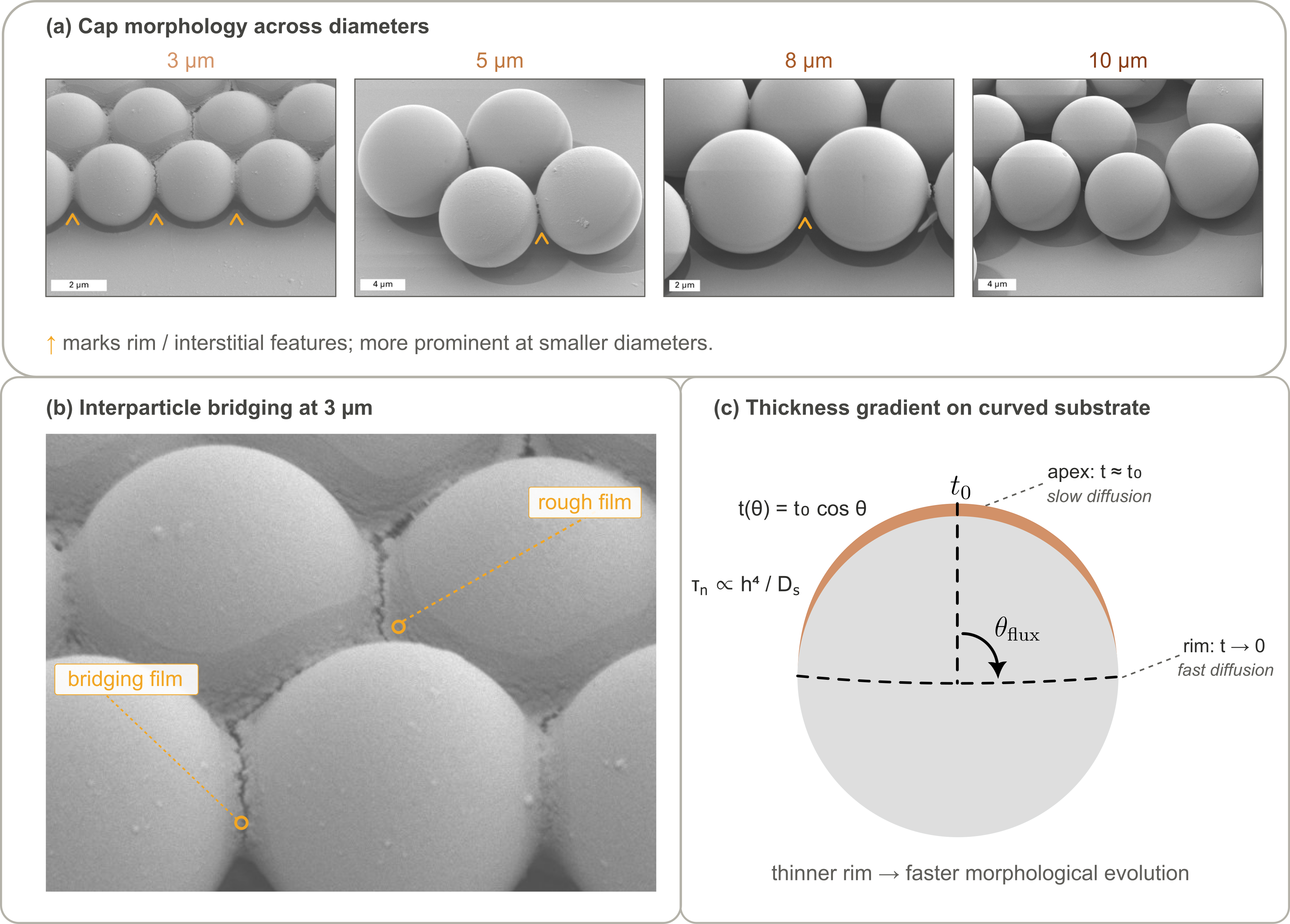}
      \caption{
    Geometric effects on cap morphology and film evolution across spherical substrates.
    (a) SEM images of FePt caps on SiO$_2$ spheres with diameters from 3 to 10~$\mu$m, highlighting systematic variations in interstitial film formation and surface features. Smaller diameters exhibit more pronounced interparticle features and local roughness.
    (b) High-magnification SEM image illustrating interparticle bridging at small diameters, where thin films form between adjacent spheres.
    (c) Schematic of thickness variation on curved substrates, where the local film thickness follows $t(\theta) = t_0 \cos\theta$, leading to thinner regions near the cap edge and thicker regions near the apex. This thickness gradient influences local diffusion kinetics and morphological evolution during annealing.
      }\label{fig:Morphology}
\end{figure}

\begin{figure}[thbp]
      \centering
\includegraphics[width=1\textwidth,page=1,trim=0.0in 0in 0in 0.0in,clip]{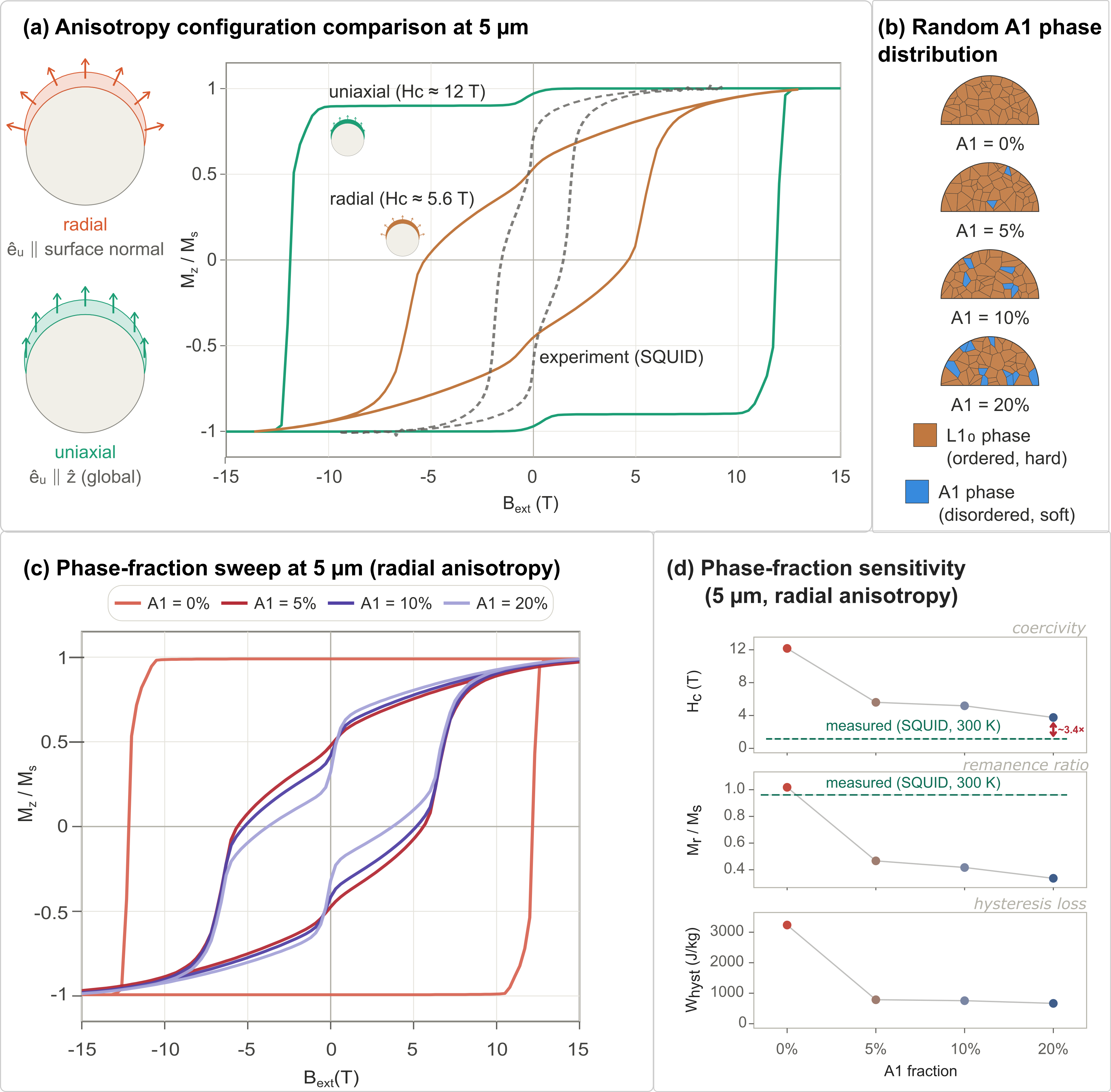}
      \caption{
      Dependence of magnetization reversal on anisotropy configuration and phase composition.
(a) Comparison of radial and uniaxial anisotropy configurations at fixed diameter (5~$\mu$m), showing that the easy-axis distribution strongly modifies the hysteresis-loop shape and coercive field. The radial configuration produces a more strongly sheared loop than the uniform uniaxial configuration.
(b) Representative spatial distributions of A1 (disordered, low anisotropy) and L1$_0$ (ordered, high anisotropy) phases for increasing A1 fraction.
(c) Simulated hysteresis loops showing progressive reductions in coercivity and remanence with increasing A1 fraction.
(d) Extracted coercivity (top), remanence ratio (middle), and hysteresis loss (bottom) as a function of A1 fraction. Measured SQUID values are overlaid on the coercivity and remanence panels for comparison. The simulated mean coercivity decreases from $12.20 \pm 0.14$~T at full L1$_0$ order to $3.80 \pm 0.14$~T at 20\% A1, whereas the measured coercivity remains lower ($1.03$--$1.21$~T). In contrast, the measured remanence ratio ($M_r/M_s = 0.917$--$0.976$) remains close to the fully ordered simulated limit, indicating that no single bulk A1 fraction reproduces both experimental quantities. Experimental values are not shown in the hysteresis-loss panel because the measured ($\pm$7~T) and simulated ($\pm$18~T) loops span different field ranges.
      }\label{fig:PhaseDependency}
\end{figure}

The simulations further showed that anisotropy configuration and phase ordering provide distinct contributions to magnetization reversal. Changing from a uniform to a radial easy-axis distribution strongly modified the hysteresis-loop shape and coercivity, demonstrating the configurational role of geometry (Figure~\ref{fig:PhaseDependency}a). Within the radial configuration, increasing the contribution of magnetically soft A1-like regions produced a strong and nonlinear reduction in coercivity and remanence (Figure~\ref{fig:PhaseDependency}b--d). The much lower coercivity measured experimentally was consistent with partial chemical ordering together with additional sources of magnetic softening not represented in the idealized simulations. SEM observations of rim roughening, grain-boundary grooving, and partial morphological evolution suggest that changes in exchange connectivity and local anisotropy may contribute to this remaining difference (Figure~\ref{fig:Morphology}).

Taken together, these results establish a hierarchy of design parameters for micrometer-scale FePt Janus caps in which size and magnetic response are decoupled. Geometry configures the anisotropy landscape; once particle diameter no longer acts as a direct tuning parameter because $R \gg \ell_{\mathrm{ex}}$, material state (the L1$_0$/A1 phase balance) becomes the primary parameter for tuning magnetization reversal, with processing-induced morphology a secondary modulation. This decoupling has a practical consequence for magnetic Janus particles used in functional colloids and microrobotic systems: particle size can be selected according to transport, payload, controllability, and biological constraints without a direct magnetic penalty, while magnetic differentiation is introduced through control of phase ordering and microstructure. More broadly, the results locate a micrometer-scale regime in which geometric configuration remains magnetically relevant even though diameter-dependent tuning has become weak, and the $\ell_{\mathrm{ex}}/R$ map provides an approximate length-scale criterion for identifying comparable regimes in other curved magnetic microsystems.

This study focuses on the static magnetic design parameters rather than on device-level performance. Magnetically driven locomotion, steering, and imaging using the same 60~nm FePt coating on micrometer-scale particles have previously been demonstrated~\cite{bozuyuk_high-performance_2022}, establishing the relevance of the physical platform. Here, we provide a complementary analysis of how particle diameter, anisotropy configuration, phase state, and processing-induced morphology contribute to its magnetic response. Dynamic actuation was not evaluated for the specific phase-ordering and morphological states investigated in this work. A direct experimental extension would therefore be to determine how controlled changes in material state translate into differences in actuation, sorting, or staged-release behavior under flow.

A complementary question follows directly from this decoupling. Because the L1$_0$ transformation and the onset of dewetting are diffusion-driven processes on a curved substrate ($t(\theta) = t_0 \cos\theta$, $\tau_n \propto h^4/D_s$), particle size may govern the phase-ordering and morphological state that is \textit{achieved} during processing, even where it does not directly tune reversal at a fixed material state. Within the 3--10~$\mu$m range studied here this indirect route did not produce a systematic magnetic size trend, but isolating it (how curvature and particle diameter set the ordering and morphology outcome of FePt on curved substrates) is a natural subject for a subsequent study.

\clearpage

\begin{figure}[thbp]
\centering
\includegraphics[width=0.80\textwidth]{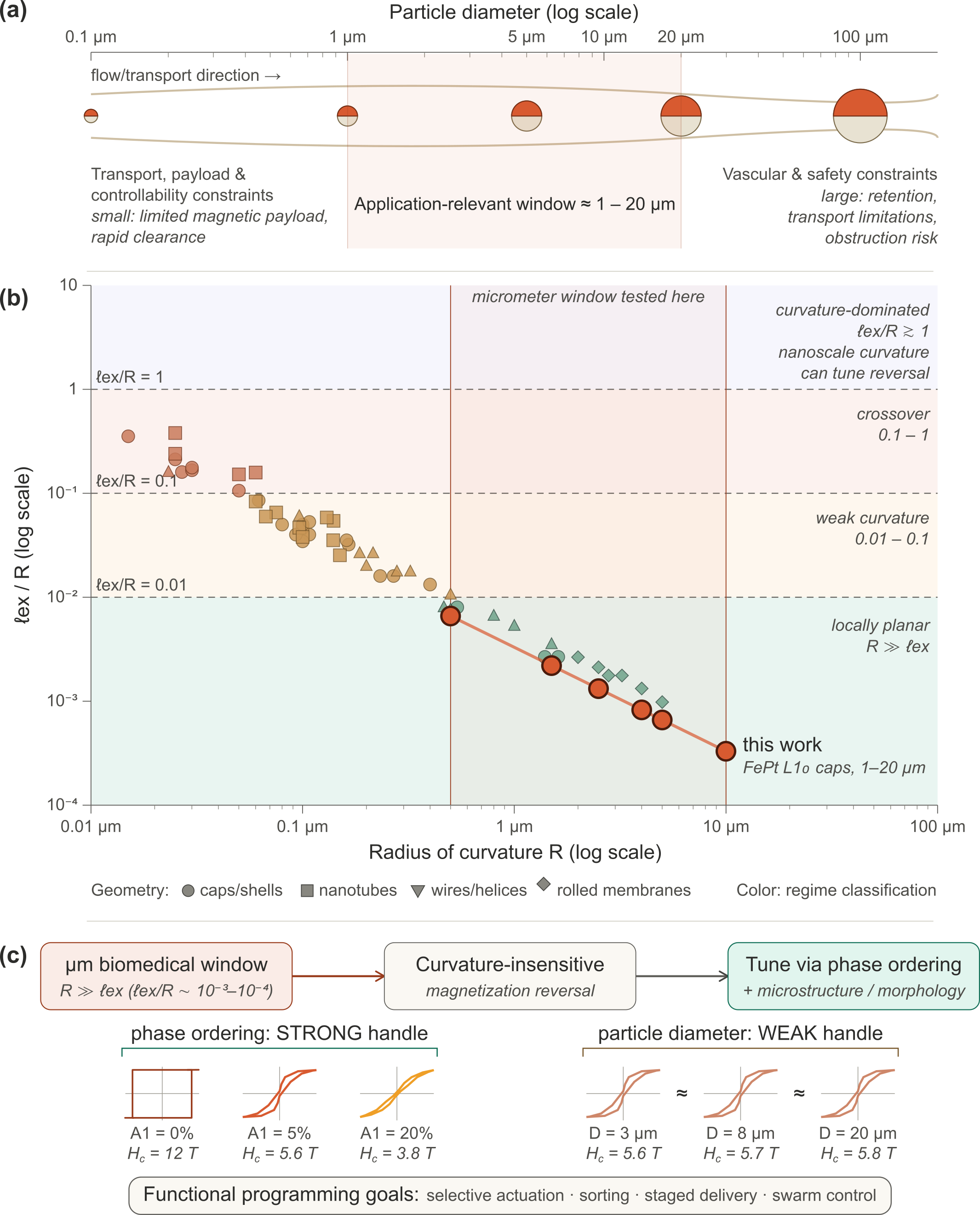}
\caption{\textbf{Curvature-regime map and design implication for FePt Janus particles.}
(a) Schematic representation of the application-relevant micrometer-scale window tested here, where particle diameter must balance magnetic payload, transport, controllability, and safety considerations.
(b) Literature-informed curvature-regime map showing the logarithmic ratio between exchange length and radius of curvature, $\ell_{\mathrm{ex}}/R$, as a function of radius of curvature $R$ for curved ferromagnetic shells, caps, nanotubes, wires, helices, and rolled membranes. Literature data points are grouped by geometry, while the FePt Janus caps investigated here occupy the locally planar regime ($R \gg \ell_{\mathrm{ex}}$). Full data point sources, extracted parameters, and references are provided in the methods and Table~S7.
(c) Design implication for magnetic Janus particles in the micrometer regime: particle diameter is a weak handle for tuning magnetization reversal, whereas phase ordering, morphology, and microstructure provide stronger routes for programming magnetic response toward envisioned functions such as selective actuation, sorting, staged delivery, and swarm-level control.}
\label{fig:RegimeMap}
\end{figure}
\newpage

\newpage

\clearpage

\clearpage

\clearpage

\clearpage

\clearpage
\begin{table}[htbp]
\centering
\caption{Comparison of repeated experimental SQUID measurements and micromagnetic simulations for FePt Janus caps. Experimental values are reported as the two repeated measurements acquired from each sample chip to convey measurement repeatability directly. Simulation values are per-diameter results from single-cap simulations with A1 fractions of 0\%, 5\%, 10\%, and 20\%. The small variation in simulated $H_c$ across diameters reflects numerical convergence rather than physical curvature dependence.}
\label{tab:SQUID_sim_comparison}
\small
\begin{tabular}{c c c c c c c}
\hline
\textbf{Diameter} & \boldmath$\mu_0 H_c^{\mathrm{exp}}$ & \boldmath$M_r/M_s^{\mathrm{exp}}$ & \boldmath$\mu_0 H_c^{0\%}$ & \boldmath$\mu_0 H_c^{5\%}$ & \boldmath$\mu_0 H_c^{10\%}$ & \boldmath$\mu_0 H_c^{20\%}$ \\
($\mu$m) & (T) & (unitless) & (T) & (T) & (T) & (T) \\
\hline
3  & 1.205, 1.127 & 0.917, 0.976 & 12.09 & 5.55 & 5.14 & 3.72 \\
5  & 1.156, 1.097 & 0.963, 0.926 & 12.17 & 5.61 & 5.18 & 3.76 \\
8  & 1.145, 1.028 & 0.954, 0.967 & 12.16 & 5.68 & 5.25 & 3.91 \\
10 & 1.154, 1.107 & 0.967, 0.958 & 12.36 & 5.67 & 5.28 & 3.78 \\
\hline
\end{tabular}
\end{table}
\clearpage

\section*{Methods}

\subsection*{Particle Monolayer Preparation and FePt Deposition}

Monodisperse SiO$_2$ particles (3--10~$\mu$m diameter, microParticles GmbH) were assembled into near-monolayers on Si substrates ($5 \times 5$~mm$^2$) \textit{via} drop-casting of diluted aqueous suspensions. Dilution factors were determined from geometric packing considerations to achieve close-packed coverage (Supporting Information). Substrates were cleaned by sequential sonication in deionized water and ethanol/acetone (1:1), followed by oxygen plasma treatment for ten minutes at 30 W to enhance wettability. A 25~$\mu$L droplet was deposited immediately after plasma activation and dried under ambient conditions for at least 12 hours.

FePt thin films were deposited by molecular beam epitaxy \textit{via} co-evaporation of Fe and Pt under high vacuum ($< 1 \cdot 10^{-8}$ mbar). Fluxes were adjusted to obtain near-equiatomic composition (50:50 $\pm$ 5 at \%), and a nominal thickness of 60 nm was monitored \textit{in situ} using a quartz crystal microbalance. Continuous substrate rotation ensured uniform azimuthal coverage.

\subsection*{Thermal Processing and Structural Characterization}

Post-deposition annealing was performed at $500^\circ$C for one hour under argon atmosphere to promote partial A1 $\rightarrow$ L1$_0$ phase transformation while limiting excessive solid-state dewetting of the FePt caps. Preliminary experiments indicated that FePt films on curved/spherical substrates exhibited the onset of diffusion-driven morphological evolution at lower temperatures than comparable planar films, particularly near the cap perimeter where the local film thickness decreases. These conditions therefore represent a compromise between chemical ordering and preservation of cap continuity during annealing. Structural and morphological characterization were carried out using field-emission scanning electron microscopy (FE-SEM, Zeiss GeminiSEM500, 2–8 kV) in both top-view and tilted configurations to assess cap continuity and curvature-dependent morphology. Elemental composition was verified by energy-dispersive X-ray spectroscopy (EDX, 25 kV, area-averaged mapping), confirming near-equiatomic FePt. Phase identification was performed by X-ray diffraction (XRD) using Cu K$\alpha$ radiation with cross-beam optics.

\subsection*{Magnetic Characterization}

Magnetic hysteresis loops were measured using a Quantum Design MPMS3 SQUID magnetometer in vibrating sample mode at 300 K with applied fields up to $\pm$7 T. Diamagnetic background contributions from the substrate were subtracted using linear fits to the high-field regions (Supporting Information), and magnetization was normalized to the saturation value. The coercive field ($H_c$), remanence ratio ($M_r/M_s$), and hysteresis loss ($W_{\mathrm{hyst}}$) were extracted from the corrected loops.

\subsection*{Micromagnetic Simulations}

Micromagnetic simulations were performed using OOMMF via the Ubermag Python interface. FePt caps were modeled as hemispherical shells with inner radius $R = d/2$ and thickness $t = 60$ nm, using material parameters representative of L1$_0$ FePt ($M_s = 1.0 \times 10^6$ A/m, $A = 1.0 \times 10^{-11}$ J/m, $K_u = 6.6 \times 10^6$ J/m$^3$). Unless otherwise stated, the easy axis was aligned along the local surface normal to represent a radial anisotropy configuration. Hysteresis loops were computed \textit{via} quasi-static energy minimization by sweeping the applied field between $\pm$18 T with a small transverse bias field to avoid metastable states. Additional numerical details and geometry implementation are provided in the Supporting Information. 

\subsection*{Literature-Based Curvature Regime Map}

To contextualize the length-scale regime investigated in this work, we compiled previously published curved ferromagnetic systems spanning radii of curvature from 0.01 to 100~$\mu$m. Spherical caps and shells were included from studies of nanosphere-supported magnetic caps and multilayers~\cite{albrecht_magnetic_2005,ulbrich_magnetization_2006,eimuller_spin-reorientation_2008}~\cite{amaladass_nanospheres_2010,amaladass_size_2007,makarov_fept_2008}, hemispherical and permalloy cap arrays~\cite{streubel_equilibrium_2012,streubel_magnetic_2012,abdelgawad_magnetic_2018,kostopoulos_nanocaps_2019}, Janus or bistable cap systems~\cite{philipp_magnetic_2021,aravind_bistability_2019,sam_size-dependent_2024,brandt_size-dependent_2013}, and spherical-shell or hemishell models~\cite{yang_intrinsic_2021,sloika_geometry_2017,kravchuk_topologically_2016,mourkas_curvature_2021,schultz_micromagnetic_2017}. Nanotube systems were included from studies reporting reversal-mode crossovers and angular switching behavior~\cite{landeros_reversal_2007,escrig_crossover_2008,bachmann_size_2009,escrig_phase_2007,albrecht_experimental_2011}, single-nanotube magnetic states and vortex configurations~\cite{wyss_imaging_2017,weber_cantilever_2012,ruffer_magnetic_2012,zimmermann_origin_2018}, and additional nanotube reversal, chirality, and spin-texture studies~\cite{proenca_crossover_2012,chen_magnetization_2010,baumgaertl_magnetization_2016,josten_curvature-mediated_2021,korber_curvilinear_2022}. Rolled membranes and curved films were included from studies of spin waves and rolled-up magnetic membranes~\cite{mendach_spin-wave_2008,bermudez_urena_fabrication_2009,balhorn_spin-wave_2010,streubel_magnetically_2012,muller_tuning_2009}, buried and single-layer rolled nanomembranes~\cite{streubel_rolled-up_2013,streubel_imaging_2014,streubel_magnetic_2014}, and parabolic stripes, curved nanowires, and domain-wall pinning in curved conduits~\cite{volkov_experimental_2019,volkov_experimental_2019-1,lewis_magnetic_2009,yershov_curvature-induced_2015,moreno_oscillatory_2017,brajuskovic_understanding_2021}. Helices, twisted ribbons, and three-dimensional curved nanowires were included from studies of magnetochiral coupling and artificial helical architectures~\cite{pylypovskyi_coupling_2015,sanz-hernandez_artificial_2020}, domain-wall automotion and three-dimensional nanohelices~\cite{skoric_domain_2022,fullerton_understanding_2024}, and geometric chirality or fractional skyrmion tubes in three-dimensional magnetic structures~\cite{xu_geometry-induced_2026,fullerton_fractional_2026,farinha_interplay_2025}. These systems were compared using the dimensionless ratio $\ell_{\mathrm{ex}}/R$ (Figure~\ref{fig:RegimeMap}b; Table~S7), where $R$ is the curvature radius and $\ell_{\mathrm{ex}}$ is the exchange length. The full curated and annotated dataset, including extracted material system, geometry, topology, regime assignment, parameter sources, and notes for each published work, is provided in Table~S7.

\section*{Acknowledgments}
This work was supported by the Alexander von Humboldt Foundation (to A.K.S.) and the Max Planck Society (to G.R.).

\noindent\textbf{Author Contributions:} N.G.V., E.S., A.K.S., and G.R. conceived the idea and designed the experiments. E.S., A.K.S., and G.R. supervised the research project. N.G.V., E.S., E.G., R.O.M.R., H.D., F.T., J.U., and A.K.S. conducted the experiments. N.G.V., E.G., R.O.M.R., A.K.S., and G.R. analyzed the data. All authors contributed to the writing and editing of the manuscript. \textbf{Competing interests:} The authors declare no competing interests. 

\noindent\textbf{Data and materials availability:} All data are available in the main text, Supporting Information, or online Edmond repository~\cite{gonzalez_vazquez_dataset_2026}. Simulation scripts, analysis notebooks, and reproducibility workflows are publicly available through the FunMaP GitHub repository (\href{https://github.com/nagova/FunMaP}{https://github.com/nagova/FunMaP}). 

\noindent\textbf{Declaration of generative AI and AI-assisted technologies in the manuscript preparation process}: During the preparation of this work, the author(s) used large language models (LLMs) to assist with the grammatical corrections of individual sentences. After using this tool/service, the author(s) reviewed and edited the content as needed and take(s) full responsibility for the content of the published article.

\noindent\textbf{Supporting Information Available:} Supplementary methods for monolayer preparation, FePt deposition, annealing, XRD, SQUID analysis, micromagnetic simulations, numerical considerations, as-deposited reference measurements, and the literature-based curvature-regime pre-published work table.

\singlespacing
\begin{footnotesize}
\bibliography{bibliography/references}
\end{footnotesize}
\bibliographystyle{achemso}

\end{document}